 \documentclass[10pt]{article}

\def\hybrid{\topmargin 0pt      \oddsidemargin 0pt
 \headheight 0pt \headsep 0pt \textheight 9in \textwidth 6.25in
\marginparwidth .875in \parskip 5pt plus 1pt \jot = 1.5ex}

\catcode`\@=11
\def\marginnote#1{}
\def\numberbysection{\@addtoreset{equation}{section}
        \def\theequation{\thesection.\arabic{equation}}}
\catcode`@=12
\relax
\numberbysection
\hybrid

\begin{document}

\title{{\Large Differential Galois groups of high order
 Fuchsian ODE's\footnote{2005 Nankai conference
on differential geometry in the honor of Professor Shiing Shen Chern.
}.}}

 \author{ N. Zenine$^\S$, S. Boukraa$^\dag$, S. Hassani$^\S$, J.-M. Maillard$%
^\ddag$}
\date{\today}
\maketitle
\centerline{$^\S$ C.R.N.A.,
Bld Frantz Fanon, BP 399, 16000 Alger, Algeria
}
\vskip .1cm
\centerline{$^\dag$ Universit\'e de Blida, Institut d'A{\'e}ronautique,
 Blida, Algeria}
\vskip .1cm \centerline{$^\ddag$ LPTMC, Universit\'e de Paris 6,} \vskip %
.1cm  
\centerline{Tour 24, $4^{eme}$ \'etage,
case 121, 4 Place Jussieu, } \vskip .1cm \centerline{ F--75252 Paris
Cedex 05, France\footnote{maillard@lptmc.jussieu.fr, maillard@lptl.jussieu.fr, 
sboukraa@wissal.dz, njzenine@yahoo.com}
}
\vskip .5cm

\begin{abstract}
We present a simple, but efficient, way to calculate
connection matrices between sets of independent local solutions, defined at
two neighboring singular points, of Fuchsian differential equations of
quite large orders, such as those found for the
 third and fourth contribution ($\chi^{(3)}$ and $\chi^{(4)}$)
 to the magnetic susceptibility of square lattice Ising model.
 We use the previous connection matrices
to get the exact explicit expressions of all the monodromy matrices 
of the Fuchsian differential equation for $\chi^{(3)}$ (and $\chi^{(4)}$)
expressed in the same basis of solutions.
These monodromy  matrices  are the generators of the differential Galois
 group of the Fuchsian
differential equations for $\chi^{(3)}$ (and $\chi^{(4)}$),
whose analysis is just sketched here.
\end{abstract} 
\vskip .2cm
\noindent {\bf AMS Classification scheme numbers}: 34M55, 47E05, 81Qxx, 32G34, 34Lxx, 34Mxx, 14Kxx
\vskip .5cm
 {\bf Key-words}:  Susceptibility of the Ising model, series expansions, 
 singular behaviour, asymptotics, Fuchsian differential equations,  
 apparent singularities, rigid local systems, differential
 Galois group, monodromy group, Euler's and Catalan's constant, 
Clausen function, polylogarithms, Riemann zeta function, multiple zeta values.

\section{Introduction}
\label{intro}

Since the work of T.T. Wu, B. M. McCoy, 
C.A. Tracy and E. Barouch~\cite{wu-mc-tr-ba-76}, 
it is known that the expansion in $n$-particle contributions to the zero field 
susceptibility of the square lattice Ising model
at temperature $T$ can be written as an infinite sum:
\begin{eqnarray}
\label{coywu}
\chi(T) = \sum_{n=1}^{\infty} \chi^{(n)}(T) 
\end{eqnarray}
of $(n-1)$-dimensional 
integrals~\cite{nappi-78,pal-tra-81,yamada-84,yamada-85,nickel-99,nickel-00},
the sum being restricted to odd (respectively even) $n$ for the high
(respectively low) temperature case.

As far as  regular singular points are concerned (physical or 
non-physical singularities in the complex plane), 
and besides the known $\, s= \pm 1$ and $\, s\, = \pm i\, $ singularities, 
B. Nickel showed~\cite{nickel-99} 
that  $\, \chi^{(2\, n+1)}$ is singular
 for the following finite values of
$\, s=sh(2\,J/kT) \,$ lying on the $\, |s\, =\, 1|$ 
{\em unit circle} ($m=k=0$ excluded):
 \begin{eqnarray}
\label{sols}
&&2 \cdot \Bigl(s \, + \, {{1} \over {s}}\Bigr) \, = \, \, \, \, 
u^k\, + \,{{1} \over {u^k}} \,
+ \, u^m\, + \,{{1} \over {u^m}} \,\nonumber \\
&&  \qquad u^{2\, n+1} \, = \, \, 1,  \qquad   \qquad  
-n \,  \,\le\,  \, m, \, \, k\,  \,\,\le \, \, n  
\end{eqnarray}
In the following we will call these singularities:  ``Nickel singularities''. 
When $n$ increases, the singularities of the higher-particle components 
of $\chi(s)$ {\em accumulate on the unit circle} $\, |s|\, = \, 1$.
The existence of such a {\em natural boundary} for the total $\,\chi(s)$, 
shows that  $\,\chi(s)$ is {\em  not D-finite} (non 
holonomic\footnote{The fact this natural boundary
may be a ``porous'' natural frontier allowing some
 analytical continuation through it is not relevant
here: one just need an infinite accumulation of 
singularities (not necessarily on a curve ...) to rule out the
D-finite character of $\, \chi$.} 
as a function of $\, s$).

A significant amount of work had been performed to
generate {\em isotropic} series coefficients
for $\chi^{(n)}$ (by B. Nickel \cite{nickel-99,nickel-00} up to order 116,
then to order 257 by 
A.J. Guttmann and W. Orrick\footnote{A.J. Guttmann
 and W. Orrick private communication.}).
More recently,  W. Orrick {\it et al.}~\cite{or-ni-gu-pe-01b},
have generated coefficients\footnote{The short-distance terms  were shown to
have the form $\, (T-T_c)^p \cdot (log|T-T_c|)^q\, $
with $\, p \ge q^2$. }
 of $\chi (s)$ up to order 323 and 646 for high and low temperature 
series in $s$,
using  some {\em non-linear Painlev\'e difference equations}
for the correlation
 functions~\cite{or-ni-gu-pe-01b,or-ni-gu-pe-01,coy-wu-80,perk-80,jim-miw-80}.
As a consequence of this  non-linear Painlev\'e difference equation,
and the remarkable associated {\em quadratic double recursion}
 on the correlation functions, the computer 
algorithm had a $\, O(N^6)$ {\em polynomial growth} 
of the calculation of the series expansion 
instead of an exponential growth
that one would expect
at first sight. However, in such a {\em non-linear}, non-holonomic,
{\em Painlev\'e-oriented} approach,
 one obtains results directly for the total susceptibility $\chi (s)$
 which do not satisfy
any linear differential equation, and thus prevents the easily 
disentangling of the contributions of the various holonomic $\chi^{(n)}$'s. 

In contrast, we consider here, a {\em strictly holonomic approach}. 
This approach~\cite{ze-bo-ha-ma-04,ze-bo-ha-ma-05,ze-bo-ha-ma-05b} enabled
 us to get 490 coefficients\footnote{We thank J. Dethridge
for writing an optimized C++ program that confirmed the Fuchsian ODE
we found for  $\, \chi^{(3)}$, providing 
hundred more coefficients all in agreement with
our Fuchsian ODE. }
of the series expansion of $\, \chi^{(3)}$ (resp. 390 
coefficients for $\, \chi^{(4)}$), from which we
 have deduced~\cite{ze-bo-ha-ma-04,ze-bo-ha-ma-05,ze-bo-ha-ma-05b,zenine4}
the {\em Fuchsian differential equation} of
order {\em seven} (resp.{\em ten}) satisfied 
by $\chi^{(3)}$ (resp. $\chi^{(4)}$). We will focus, here, on the 
{\em differential Galois group} of these order seven and ten Fuchsian 
ODE's.

\section{The Fuchsian differential equations satisfied by $\,\tilde{\protect%
\chi}^{(3)}(w)$ and $\tilde{\protect\chi}^{(4)}(w)$}
\label{ODE}

Similarly to Nickel's papers~\cite{nickel-99,nickel-00}, we start using the
multiple integral form of the $\,\chi ^{(n)}$'s, or more precisely of some
normalized expressions $\,\tilde{\chi}^{(n)}$:
\begin{eqnarray}
\chi ^{(n)}(s) &=&\,S_{\pm }\;\;\tilde{\chi}^{(n)}(s),\qquad \quad
n=3,\,4,\,\cdots   \label{KHIn} \\
S_{+} &=&\frac{{\ (1-s^{4})^{1/4}}}{{s}}, \qquad T>T_{C}\quad \quad (n%
\quad odd)  \nonumber \\
S_{-} &=&{(1-s^{-4})^{1/4}}, \qquad T<T_{C}\quad \quad (n \quad even)
\nonumber
\end{eqnarray}%
where:
\begin{equation}
\label{KHInTILDA}
\tilde{\chi}^{(n)}(w)\, =\,\,\int d^{n}V\;\quad \left(
\prod_{i=1}^{n}\tilde{y}_{i}\right) \;\cdot R^{(n)}\cdot H^{(n)}
\end{equation}%
with (each angle ${\phi _{i}}$ varying from $0$ to $2\pi $):
\begin{eqnarray} 
\label{VRH}
&&d^{n}V =\;\quad \prod_{i=1}^{n-1}\frac{{d\phi _{i}}}{2\pi }\quad 
with \quad \sum_{i=1}^{n}{\phi _{i}=0}, \quad \quad  
R^{(n)} =\, \frac{1+\prod_{i=1}^{n}\tilde{x}_{i}}
{1-\prod_{i=1}^{n}\tilde{x}_{i}}  \nonumber \\
&&H^{(n)} =\, \prod_{i<j}4\frac{\tilde{x}_{i}\tilde{x}_{j}}{\left( 1-%
\tilde{x}_{i}\tilde{x}_{j}\right) ^{2}} 
\cdot \sin^{2}\left( \frac{{\phi _{i}-\phi_{j}}}{2}\right)  
\end{eqnarray}%
Instead of the usual\cite{nickel-99,nickel-00} variable $s$, we found it
more suitable to use $w\,=\,{\frac{{1}}{{2}}}s/(1+s^{2})\,$ which has, by
construction, Kramers-Wannier duality invariance ($s\leftrightarrow 1/s$)
and thus allows us to deal with both limits (high and low temperature, small
and large $s$) on an equal footing \cite%
{ze-bo-ha-ma-04,ze-bo-ha-ma-05,ze-bo-ha-ma-05b}. The quantities $\tilde{x}%
_{j}$ and $\tilde{y}_{j}$ can be written in the following form \cite%
{ze-bo-ha-ma-04,ze-bo-ha-ma-05,ze-bo-ha-ma-05b}:   
\begin{eqnarray}
\label{varxy}
\tilde{x}_{j} &=&\,\,{\frac{2w}{1-2w\cos {\phi _{j}}+\sqrt{(1-2w\cos {\phi
_{j}})^{2}-4w^{2}}}},  \nonumber \\
\tilde{y}_{j} &=&\,\,{\frac{2w}{\sqrt{(1-2w\cos {\phi _{j}})^{2}-4w^{2}}}}.%
\end{eqnarray}%
It is straightforward to see that $\,\tilde{\chi}^{(n)}$ is \emph{only a
function of the variable} $w$. From now on, we thus focus on $\,\tilde{\chi}%
^{(n)}$ seen as a function of the well-suited variable $\,w$ instead of $\,s$
\cite{nickel-99,nickel-00}. One may expand the integrand in (\ref{KHInTILDA}%
) in this variable $\,w$, and integrate the angular part.

We do not recall, here, the concepts, tricks and tools that have been
necessary to generate very large series expansion for $\,\tilde{\chi}%
^{(3)}(w)$ and $\,\tilde{\chi}^{(4)}(w)$ with a \emph{polynomial growth} of
the calculations~\cite{ze-bo-ha-ma-04,ze-bo-ha-ma-05,ze-bo-ha-ma-05b}.

Given the expansion of $\,\tilde{\chi}^{(3)}(w)$ up to $\,w^{490}$, the next
step amounts to encoding all the numbers in this long series into a \emph{%
linear} differential equation. Note that such an equation should exist
though its order is \emph{unknown}\footnote{%
A lower bound for the order of this linear differential equation would be
extremely useful : such a lower bound \emph{does not} exist at the present
moment.}. Let us say that, using a dedicated program for searching%
\footnote{%
Note that we, first, actually found an order twelve Fuchsian ODE and, then,
we reduced it (by factorization of the differential operator) to a seventh
order operator. This order twelve differential equation requires much less
coefficients in the series expansion to be guessed than the order seven
Fuchsian ODE we describe here ! It is \emph{actually easier to find the
order twelve} differential equation than the order seven ODE !!} for such a
finite order linear differential equation with polynomial coefficients in $w$%
, we succeeded finally in finding the following linear differential equation
of order \emph{seven} satisfied by the $490$ terms we have calculated for $\,%
\tilde{\chi}^{(3)}$: 
\begin{eqnarray}
&&\sum_{n=0}^{7}\,a_{n}\cdot {\frac{{d^{n}}}{{dw^{n}}}}F(w)\,\,=\,\,\,\,\,0%
\quad \quad \quad \quad \quad \hbox{with:}  \label{fuchs} \\
&&a_{n}\,=\,\,w^{n}\cdot (1-4\,w)^{\theta (n-2)}\,(1+4\,w)^{\theta
(n-4)}\,P_{n}(w),\quad n=6,\,5,\cdots ,\,0  \nonumber \\
&&\hbox{where:}\quad \quad \quad \theta (m)\,=\,sup(m,\,0),\quad \quad \quad %
\hbox{and:}\quad \quad \quad a_{7}\,=\,\,  \nonumber \\
&&w^{7}\cdot \left( 1-w\right) \,\left( 1+2\,w\right) \,{\left(
1-4\,w\right) }^{5}\,{\left( 1+4\,w\right) }^{3}\,\left(
1+3\,w+4\,w^{2}\right) \,P_{7}(w)  \nonumber
\end{eqnarray}%
where $P_{7}(w),P_{6}(w)$ $\cdots $, $P_{0}(w)$ are polynomials of degree
respectively 28, 34, 36, 38, 39, 40, 40 and 36 in $\,w$~\cite{ze-bo-ha-ma-04}%
.

Furthermore, besides the known singularities (\ref{sols}) mentioned above,
we remark the occurrence of the roots of the polynomial $\,P_{7}$ of degree
28 in $w$, and the two \emph{quadratic numbers} 
roots of $\,1+3\,w+4\,w^{2}\,=\,0$
which are not~\cite{Nickel-05} Nickel singularities (they are 
not of the form (\ref{sols})). The two quadratic
numbers \emph{are not} on the $s$-unit circle : $|s|=\,\sqrt{2}$ and $%
|s|=\,1/\sqrt{2}$. These quadratic
numbers do not occur in the ``physical solution'' $\, \chi^{(3)}$.
 For $P_{7}$, near any of its roots, all the local
solutions carry \emph{no logarithmic terms} and \emph{are analytical} since
the exponents \emph{are all positive integers}. The roots of $P_{7}$ are
thus \emph{apparent singularities} \cite{ince-56,Forsyth} of the Fuchsian
equation (\ref{fuchs}).

The order seven linear differential operator $L_{7}$ associated with the
differential equation satisfied by $\tilde{\chi}^{(3)}$ has the following
factorization properties \cite{ze-bo-ha-ma-04,ze-bo-ha-ma-05,zenine4}:%
\begin{eqnarray}
\label{FactorL7}
L_{7} \, =\, \, L_{1}\oplus L_{6}, \qquad \quad
 L_{6} \, = \, \,Y_{3}\cdot Z_{2}\cdot N_{1} 
\end{eqnarray}%
where\footnote{%
For the notations see \cite{ze-bo-ha-ma-04,ze-bo-ha-ma-05,zenine4} for $%
\tilde{\chi}^{(3)}$, and \cite{ze-bo-ha-ma-05b,zenine4} for $\tilde{\chi}%
^{(4)}$.} $L_{1}$ is a first order differential 
operator which has the first contribution
to the magnetic susceptibility, namely $\tilde{%
\chi}^{(1)}=\, 2w/(1-4w)$, as solution.

In the same way, we found that the order ten linear differential operator $%
L_{10}$, associated with the differential equation satisfied by $\tilde{\chi}%
^{(4)}$, has the following factorization properties \cite%
{ze-bo-ha-ma-05b,zenine4}:
\begin{eqnarray}
\label{FactorL10}
L_{10} \,=\,\, N_{0}\oplus L_{8}, \qquad \qquad 
L_{8} \, =\,\, M_{2} \cdot G(L)  
\end{eqnarray}%
where $N_{0}$ is an order two differential operator 
which has the second contribution
to the magnetic susceptibility, $\tilde{\chi}%
^{(2)}$ as solution and where $\, G(L)$ is an order four differential  operator
that can be factorized in a product of four 
order one differential operators~\cite{ze-bo-ha-ma-05b}. 
\vskip .2cm

\section{Differential Galois group}
\label{galois}

A fundamental concept to understand
 (the symmetries, the solutions of) these
exact Fuchsian differential equations is the so-called {\em differential
Galois group}~\cite{Moscou,Put,BermSing,Berm,Weil2}, which
requires the computation of all the {\em  monodromy matrices}
associated with each (non apparent) regular singular point, these matrices
being considered {\em in the same basis}\footnote{These monodromy matrices
are the generators of the {\em monodromy group} which identifies with the
differential Galois group when there are no irregular singularities,
and, thus, no Stokes matrices~\cite{Sto}.}.  
Differential  Galois groups have been calculated
for simple enough second order, or even third order, ODE's. However, finding 
the differential Galois group of such {\em higher order}
Fuchsian differential equations (order 
seven for $\chi^{(3)}$, order ten for $\chi^{(4)}$)
 with eight regular singular points (for $\chi^{(3)}$) is not an easy task. 
Along this side a first step amounts to seeing that 
the corresponding (order seven, ten) differential operators
{\em do factorize} in smaller order 
differential operators, as a consequence of 
some rational and algebraic solutions and other
 singled out solutions~\cite{zenine4}. 
These factorizations yield a particular block-matrix form of the
 monodromy matrices~\cite{zenine4}.
The calculation of local monodromy matrices in some ``well-suited'' local
(Frobenius series solution) bases is easy to perform, however
the calculation of the so-called {\em connection matrices} corresponding to 
 the ``matching'' of the various well-suited 
local bases associated with the various 
regular singularities is a hard {\em non-local}
 problem. Of course from the knowledge of all these 
 connection matrices one can immediately write 
the monodromy matrices in a {\em unique} 
 basis of solutions\cite{zenine4}. 

From exact Fuchsian ODE's one can calculate very large series expansions 
for these (well-suited local Frobenius) solutions, 
sufficiently large that the evaluation
of these series far away from any regular singularity 
can be performed\footnote{Within the radius 
of convergence of these series.} with a very large accuracy
(400, 800, 1000 digits ...). As far as $\, \chi^{(3)}$ 
is concerned one can reduce~\cite{zenine4} 
the calculation of these connection and monodromy
 matrices, to the $\, 6 \times 6$ matrices of
an order six~\cite{zenine4}  differential operator $\, L_6$
 appearing in the decomposition (\ref{FactorL7}). Connecting 
various sets of Frobenius series-solutions
well-suited to the various sets of regular singular points
 amounts to solving a linear system of 36 unknowns
(the entries of the connection matrix). We have 
obtained these entries in floating point form with a
very large number of digits (400, 800, 1000, ...).
We have, then, been able to actually ``recognize'' these entries obtained
in floating form with a large number of digits~\cite{zenine4}.

 In particular it is shown in~\cite{zenine4}
 that the {\em connection 
matrix} between the singularity points $0$ and  $1/4$ (matching the well-suited
local series-basis near $\, w=0$ and the well-suited
local series-basis near $\, w=1/4$) is a matrix  where
 the entries are expressions in terms of 
 $\sqrt{3}$, $\pi$, $1/\pi$, $1/\pi^2$, ... 
and a (transcendental) constant  $\, I_3^{+}$ 
introduced in equation (7.12) of~\cite{wu-mc-tr-ba-76}:
\begin{eqnarray} 
\label{Wu}
&& {{1} \over {2\, \pi^2}} \cdot 
\int_{1}^{ \infty}\int_{1}^{\infty} \int_{1}^{ \infty} 
dy_1\, dy_2\, dy_3 \Bigl({{y_2^2\, -1 }
 \over {(y_1^2\, -1)\,(y_3^2\, -1)  }}\Bigr)^{1/2} \cdot 
Y^2\, = \, \, \,\nonumber \\
&&\, = \, \, \, .000814462565662504439391217128562721997861158118508 \cdots 
\nonumber \\
&& Y \, \, = \, \,\,\, {{ y_1\, -y_3} \over { 
(y_1\, +y_2)\,(y_2\, +y_3)\,(y_1\, +y_2\, +y_3)  }}. \nonumber
\end{eqnarray} 
This transcendental constant can actually be written in term of the
 {\em Clausen function} $\,Cl_2$ :
\begin{eqnarray}
\label{I3plus}
I_3^{+} \,\,  = \, \, {{1} \over {2 \pi^2 }}\cdot \Bigl( {{\pi^2} \over {3}} \,
 +2 \, -3 \sqrt{3}\cdot  Cl_2({{\pi} \over {3}})  \Bigr)
\end{eqnarray}
where  $\,Cl_2$ denotes  the
 Clausen function :
\begin{eqnarray}
\label{Claus}
   Cl_2(\theta) \, = \, \, 
\sum^{\infty}_{n=1}  \, {{\sin(n\,\theta) } \over {n^2 }} \nonumber
\end{eqnarray}
This constant   $\,  I_3^{+} $ {\em can also be
 written in terms of dilogarithms,
 polygamma functions or Barnes G-functions} : 
\begin{eqnarray}
&&I_3^{+} \, -(1/6\,  +{\pi }^{-2})\, = \, \, \, 
 -{{ 3\, \sqrt {3} } \over {2 \, \pi^2}} 
\cdot {\it Im} \left( {\it dilog} \left( 1/2-1/2\,i\sqrt {3} \right)  \right)
\nonumber \\
&&\, = \, \, {{1} \over {16 \, \pi^2}}
 \cdot \Bigl( \Psi \left( 1,2/3 \right) +\Psi \left( 1,5/6 \right)
-\Psi \left( 1,1/6 \right) -\Psi \left( 1,1/3 \right) \Bigr) \nonumber \\
&&\, = \, \,\, -{{ \sqrt {3} } \over {2 \, \pi}}
 \cdot \Bigl( \ln  \left( 2 \right)\, +  \ln \left( \pi  \right)\,
 -6 \, \ln  \left( {\frac {G \left( 7/6 \right) }
{G \left( 5/6 \right) }} \right) 
\Bigr) \nonumber
\end{eqnarray}

The $ \, 6 \times 6$ connection matrix $\, C(0,1/4)$ 
for the order six differential operator $L_6$ 
 matching the  Frobenius series-solutions
around $\, w=0$ and the ones around $\, w=1/4$,
 reads:
\begin{eqnarray}
\label{C014}
&& C(0,\, 1/4)\, = \, \, \\
&& \left[ \begin {array}{cccccc} 
1&0&0&0&0&0\\
\noalign{\medskip}
1&0&-{\frac {9 \sqrt {3}}{64 \pi}}&0&0&0\\
\noalign{\medskip}
0 &-{\frac {3 \pi \sqrt {3}}{32}} &0&0&0&0\\
\noalign{\medskip}
5&{\frac {1}{3}}-2 \cdot I_3^{+}&{\frac {3 \sqrt {3}}
{64 \pi}}&0&0&{ \frac {1}{16 \pi^2} }\\
\noalign{\medskip}
-{\frac{5}{4}}&-{\frac {3 \pi \sqrt {3}}{32}}
 &{\frac {45 \sqrt {3}}{256 \pi}}&0&{\frac{1}{32}}&0\\
\noalign{\medskip}
{\frac{29}{16}}-{\frac{2 \pi^2}{3}}&{\frac {15 \pi \sqrt {3}}{64}}
&-{\frac {225 \sqrt {3}}{1024 \pi}}
 -{\frac {3 \pi \sqrt {3}}{64}}&{\frac {\pi^2}{64}}&0&0
\end {array} \right]
\nonumber
\end{eqnarray}
Not surprisingly\footnote{One can expect the
 entries of the connection matrices to be {\em evaluations}
of (generalizations of) hypergeometric 
functions, or solutions of Fuchsian differential equations. 
} a lot of $\, \pi$'s ``pop out'' in the 
entries of these connection matrices. We will 
keep track of the $\, \pi$'s occurring in the 
entries of connection matrices through
the introduction of the variable $\, \alpha \, = \, 2\, i\, \pi$. 

The local monodromy matrices can easily be 
calculated~\cite{zenine4} since they correspond,
mostly, to ``logarithmic monodromies'' and
 will be deduced from simple calculations
using the fact that each logarithm (or
 power of a logarithm) occurring in a (Frobenius series) solution,
 is simply changed as follows : $\, \ln(w) \, $
$\rightarrow \, \, \ln(w) \, + \, \, \Omega$, where $\, \Omega$
will denote in the following $\, 2 \, i \, \pi$. 
From the local monodromy matrix $\, Loc(\Omega)$,
expressed in the 
 $\, w\, = \, 1/4$ well-suited local series-basis,
 and from the connection matrix (\ref{C014}),
 the monodromy matrix around 
 $\, w\, = \, 1/4$,  expressed 
in terms of the $\, (w=0)$-well suited basis reads~\cite{zenine4}:
\begin{eqnarray}
\label{MalphaOmega}
&&24 \, \alpha^4 \cdot M_{w=0}(1/4)(\alpha, \, \Omega)\,\, = \, \,  
\left[ \begin {array}{cc} 
{\bf A}&{\bf 0}\\
\noalign{\medskip}
{\bf B}&{\bf C}
\end {array} \right]  \\
&&\,\qquad \quad \quad  = \, \, C(0, \, 1/4) \cdot Loc(\Omega)
\cdot C(0, \, 1/4)^{-1} \nonumber
\end{eqnarray}
where
$
\left[ \begin {array}{c} 
{\bf A}\\
\noalign{\medskip}
{\bf B}
\end {array} \right] 
$ 
and
$
\left[ \begin {array}{c} 
{\bf C}
\end {array} \right]
$ read respectively:
\begin{eqnarray}
 \left[ \begin {array}{ccc} 
-24\,{\alpha}^{4}&0&0
\\
\noalign{\medskip}
-48\,{\alpha}^{4}&24\,{\alpha}^{4}&-144\,{\alpha}^{2}\Omega
\\
\noalign{\medskip}
0&0&24\,{\alpha}^{4}
\\
\noalign{\medskip}
-48\, \rho_1 
&32\,\Omega\, \rho_2 
&48\,\Omega\, \, (9\,{\alpha}^{2}+80\,\Omega) 
\\
\noalign{\medskip}
12\,{\alpha}^{2} \rho_3 
&4 \, (75-4\,{\alpha}^{2})\,{\alpha}^{2}\Omega 
&-300\,{\alpha}^{2}\Omega
\\
\noalign{\medskip}
- \, (87+8\,{\alpha}^{2}){\alpha}^{4} 
&0&3 \, (4\,{\alpha}^{2} -75)\, {\alpha}^{2}\Omega
\end {array}
 \right],
\nonumber
\end{eqnarray}
with $\, \rho_1\, = \,  5\,{\alpha}^{4}
+8\,{\Omega}^{2}+8\,{\Omega}^{2}{\alpha}^{2} $, 
$\, \rho_2\, = \, \,4\,\Omega\,{\alpha}^{2} -75\,\Omega-15\,{\alpha}^{2}  $ and
 $\, \rho_3\, = \, 5\,{\alpha}^{2}+4\,\Omega+4\,\Omega\,{\alpha}^{2} $, 
and:
\begin{eqnarray}
C \, = \, \, \,  \left[ \begin {array}{ccc} 
24\,{\alpha}^{4}&-384\,{\alpha}^{2}\Omega&1536\,{\Omega}^{2}
\\
\noalign{\medskip}
0&24\,{\alpha}^{4}
&-192\,{\alpha}^{2}\Omega
\\
\noalign{\medskip}
0&0&24\,{\alpha}^{4}
\end {array}
 \right]
\nonumber
\end{eqnarray}
Note that the transcendental constant 
$\, I_3^{+}$ {\em has disappeared} in the 
final exact expression of (\ref{MalphaOmega}) {\em which actually depends
only on} $\, \alpha$ and $\, \Omega$.
This $\, (\alpha, \,\Omega)$ way of 
writing the monodromy matrix  (\ref{MalphaOmega})
enables to get straightforwardly the $\, N-$th power of (\ref{MalphaOmega}):
\begin{eqnarray}
\label{power}
 M_{w=0}(1/4)(\alpha, \, \Omega)^N\,\, =
 \, \,\,   M_{w=0}(1/4)(\alpha, \, N\cdot \Omega)
\end{eqnarray}

Let us introduce the following  choice  of ordering of the eight
singularities, namely $\,\infty , \, \, 1, \, \, 1/4, \, \, w_1,$
$ \, \, -1/2, \, \, -1/4, \, \, 0, \, \,w_2$ 
($w_1\, = \, (-3\, +i \, \sqrt(7))/8$ and $\, w_2\, =\, w_1^{*}$ 
are the two quadratic number 
roots of $\, 1+3\, w\, + 4\, w^2\, = \, 0$), the first monodromy matrix
$\, M_1$ is, thus, the monodromy matrix $\, M_{w=0}(\infty) $
(see (\ref{MalphaOmega})) at infinity with $\, \alpha \, = \,$
$ \Omega\, = \, 2i \, \pi$, $\,\,{\cal M}(\infty)$, 
the second monodromy $\, M_2$
matrix  being the monodromy matrix at $\, w=1$, $\,{\cal M}(1)$, ...
This is actually the particular choice  of ordering of the eight
singularities, such that a
 product of monodromy matrices {\em is 
equal to the identity matrix}\footnote{Of course, from 
this relation, one also has seven other relations
deduced by cyclical permutations.}:
\begin{eqnarray}
\label{relid}
&& M_1 \cdot  M_2 \cdot  M_3 
\cdot  M_4 \cdot  M_4 
  \cdot  M_6   \cdot  M_7 \cdot  M_8 \, = \, \,\, {\bf Id} \nonumber  \\
&&\, \quad  = \, \,\, {\cal M}(\infty) \cdot {\cal M}(1) \cdot {\cal M}(1/4) 
\cdot {\cal M}(w_1)   \\
&&\qquad \times {\cal M}(-1/2)
\cdot {\cal M}(-1/4) \cdot {\cal M}(0) \cdot {\cal M}(w_2) \nonumber 
\end{eqnarray}
It is important to note that relation (\ref{relid}) 
{\em is not verified}
by the $(\alpha, \, \Omega)$ extension (like (\ref{MalphaOmega})) 
of the monodromy matrices $\, M_i$. If one considers relation
(\ref{relid}) for the $(\alpha, \, \Omega)$ extensions of the $\, M_i$'s, one 
will find that (\ref{relid}) is 
satisfied {\em only when} $\, \alpha$ {\em is equal
 to} $\, \Omega$,  but
 (of course\footnote{A matrix identity like (\ref{relid}) yields a set of 
polynomial (with integer coefficients) relations on 
$\, \Omega\, = \, \, 2 \, i \, \pi$. The number $\, \pi$ being transcendental
it is not the solution of a polynomial with integer coefficients. These 
polynomial relations have, thus, to
 be {\em polynomial identities valid for any} $\, \Omega$.})
 this $\, \alpha \, = \, \Omega$ matrix
 identity {\em is verified for any value of}
 $\, \Omega$, not necessarily equal
to $\, 2 \, i \, \pi$. 

\subsection{Mutatis mutandis for  $\, \chi^{(4)}$}
\label{mutatis}
Similarly to  $\, \chi^{(3)}$ the differential operator for $\, \chi^{(4)}$
presents remarkable factorizations that yield
 a particular block-matrix form of the
 monodromy matrices~\cite{zenine4}.
Similarly, again, one can consider the 
(Frobenius series) solutions of the differential operator 
associated with $\, \chi^{(4)}$ around $x\, =\, 4\, w^2\, =0$ and around the 
ferromagnetic (and antiferromagnetic) critical point $\, x=1$ respectively.
Again the corresponding {\em connection 
matrix} (matching the solutions around the singularity
 points $x=0$ and the ones around the singularity point $x\, =\, 4\, w^2\,=1$)
have entries which  are expressions in terms
 of $\pi^2$, rational numbers but also 
of constants like constant  $\, I_4^{-}$ introduced in~\cite{wu-mc-tr-ba-76}
which can actually be written in term of the {\em Riemann zeta
function}, as follows :
\begin{eqnarray}
\label{I4moins}
I_4^{-} \, = \, \,{{1} \over {16 \pi^3 }}
\cdot \Bigl( {{4\,\pi^2} \over {9}} \, -{{1} \over {6}} 
 -{{7} \over {2}}\cdot \zeta(3)
 \Bigr) 
\end{eqnarray}
The derivation of the two results (\ref{I3plus}), (\ref{I4moins}) 
for the two transcendental constants $\,I_3^{+} $ and   $\,I_4^{-} $
has never been published\footnote{We thank C. A. Tracy for 
pointing out the existence of these
 two results (\ref{I3plus}), (\ref{I4moins}) 
and reference~\cite{Tracy}.} but 
these results appeared in a conference proceedings~\cite{Tracy}.
We have actually checked that $\, I_3^{+}$
and  $\, I_4^{-}$ we got in our 
calculations of connection matrices displayed in~\cite{zenine5}
 as floating numbers with respectively 421 digits and 431 digits accuracy,
{\em are actually in agreement} with the previous two formula. 
These two results (\ref{I3plus}), (\ref{I4moins})  provide
 a clear answer to the question  of how ``complicated
and transcendental'' some of the constants occurring in the entries of the
connection matrices can be. These 
two remarkable exact formulas (\ref{I3plus}), (\ref{I4moins})
are not totally surprising when one recalls the {\em deep link between 
 zeta functions, polylogarithms and 
hypergeometric series}~\cite{Tanguy,Tanguy2,Tanguy3,Tanguy4}.
Along this line, and keeping in mind that we see all our Ising susceptibility
calculations as a ``laboratory'' for other 
more general problems (Feynman diagrams, ...),
we should also recall the various papers 
of D. J. Broadhurst~\cite{Broadhurst,Broadhurst1,Broadhurst2,Broadhurst3}
where $ \, Cl_2({{\pi} \over {3}})$ and $ \,\zeta(3) $ actually
 occur in a Feynman-diagram-hypergeometric-polylogarithm-zeta
framework (see for instance equation (163) in~\cite{Broadhurst}).

Similarly to the previous results for
 $\, \tilde{\chi}^{(3)}$ the monodromy matrices written in 
the {\em same} basis of solution, deduced from the connection matrices and the 
local monodromy matrices are such that a product in a certain order of them
is the identity matrix. Denoting by $M_{x=0}(0)$, 
$M_{x=0}(1)$, $M_{x=0}(4)$ and $M_{x=0}(\infty)$
the monodromy matrices expressed in the same $x=0$ well-suited basis,
 one obtains:
\begin{eqnarray}
M_{x=0}(\infty) \cdot M_{x=0}(4) \cdot M_{x=0}(1) \cdot M_{x=0}(0)
 \, =\, {\bf Id}
\end{eqnarray}
This matrix identity is valid irrespective of the ``not yet guessed''
constants~\cite{zenine4}.

\section{Conclusion}
\label{conclu}

The high order Fuchsian equations we have sketched here 
present many interesting mathematical properties close to the ones of
the so-called {\em rigid local
 systems}~\cite{darmon}, these rigid local systems
exhibiting remarkable {\em geometrical interpretations}~\cite{Katz} as periods 
of some algebraic varieties. This ``rigidity\footnote{Let us recall
that hypergeometric functions are totally rigid.}''
emerges through the log-singularities of the solutions of these Fuchsian ODE's:
the powers of the logarithms of these 
solutions are ``smaller'' than one could expect at first sight.
It is worth noting that almost all these
 mathematical structures, or singled-out properties,
we sketched here, or in previous 
publications~\cite{ze-bo-ha-ma-04,ze-bo-ha-ma-05,ze-bo-ha-ma-05b,zenine4},
are far from being specific of the
 two-dimensional Ising model : they also occur on 
many problems of lattice statistical
 mechanics or, even\footnote{The wronskian
of the corresponding differential equation 
in~\cite{Three} is also rational, the
associated differential operator factorizes in a way totally 
similar to the Fuchsian ODE's for $\, \chi^{(3)}$ and $\, \chi^{(4)}$, large 
polynomial corresponding to apparent singularities
also occur, ...},  as 
A. J. Guttmann and I. Jensen saw it recently, 
on enumerative combinatorics problems like, for instance, the 
generating function of the three-choice polygon~\cite{Three}. 

We have also seen in some of our 
calculations~\cite{ze-bo-ha-ma-05,ze-bo-ha-ma-05b} a clear occurence of 
hypergeometric functions, hypergeometric series 
and in some of our calculations (not displayed here)
generalizations of hypergeometric functions to {\em several
 complex variables}: Appel functions~\cite{Cabral},
 Kamp\'e de F\'eriet, Lauricella-like functions, 
polylogarithms~\cite{Broadhurst,Broadhurst1,Broadhurst2,Broadhurst3}, Riemann 
zeta functions, multiple zeta values, ...  
The occurence of Riemann 
zeta function or dilogarithms  
in the two remarkable exact formulas (\ref{I3plus}), (\ref{I4moins})
is not totally surprising when one recalls the {\em deep link between 
 zeta functions, polylogarithms and 
hypergeometric series}~\cite{Tanguy,Tanguy2,Tanguy3,Tanguy4}.

We think that such ``collisions'' of concepts and structures
of different domains of mathematics (differential geometry, number theory, ...)
are {\em not} a consequence of 
the free-fermion character of the Ising model, and that
similar ``convergence'' should also be encountered on more complicated 
Yang-Baxter integrable models\footnote{The
 comparison of the Riemann zeta functions
equations obtained for the
XXX quantum spin chain~\cite{Korepin} with 
the evaluations of central binomial 
 in~\cite{Broadhurst1} 
provides a strong indication in favor of similar structures
on non-free-fermion Yang-Baxter integrable models.}, the two-dimensional 
Ising model first ``popping out'' as a consequence of 
its simplicity. 
In a specific differential framework some of 
these interesting mathematical properties
can clearly be seen in the analyzis of the 
differential Galois group of these Fuchsian equations.

We have underlined the fact that, beyond a general analyzis of 
the differential Galois group~\cite{Moscou}, one can actually 
find the exact expressions of the {\em non-local}
connection matrices from  very simple matching 
of series calculations, and deduce, even 
for such {\em high order} Fuchsian ODE's, 
{\em explicit representations} of all the monodromy matrices 
in the {\em same} (non-local) basis of solutions,
{\em providing an effective way} of 
writing explicit representations of all the elements 
of the monodromy group. The remarkable form,
 structures and properties (see (\ref{C014}), 
(\ref{MalphaOmega}), (\ref{power}), (\ref{relid}))
of the  monodromy matrices in the {\em same} (non-local)
 basis of solutions is something one could not suspect
at first sight from the general description of the  differential Galois group.
 
\hskip 2cm

\textbf{Acknowledgments}
 We thank Prof. Mo-Lin Ge 
for allowing one of us (JMM) to give a talk on these
results at the 2005 Nankai conference and for 
his kind hospitality in the Nankai University.
One of us (JMM) also thank Prof. F. Y. Wu
for his interest in our
 work, his extremely valuable advice  and for 
his guidance in the past. 
One of us (JMM) thanks  Prof. Chengming Bai for help
throughout his stay in Nankai University, where part of this
paper was written. 
 We thank Jacques-Arthur Weil for valuable
comments on differential Galois group, and connection matrices.
We would like to thank B. Nickel for an inspired comment
on solution $\, S_3$. We thank C. A. Tracy for pointing out
the results of reference~\cite{Tracy}. We thank J. Dethridge for a great C++ 
program that gave an additional and very strong
 confirmation of our $\, \chi^{(3)}$
results.  One of us (JMM) thanks R.J. Baxter 
for kind hospitality and discussions when visiting the ANU in Canberra.
We would like to thank A. J. Guttmann, I. Jensen and W. Orrick
for a set of useful comments on the singularity behaviour 
of physical solutions.
(S. B) and (S. H) acknowledge partial support from PNR3-Algeria.

\end{document}